\documentclass[conference, a4paper]{IEEEtran}

\usepackage{amsmath}
\usepackage{balance}
\usepackage[T1]{fontenc}
\usepackage[pdftex]{graphicx}
\usepackage{url}

\graphicspath{{figures/}}
\DeclareGraphicsExtensions{.pdf,.png,.jpg}

\usepackage[absolute]{textpos}
\usepackage[hidelinks]{hyperref}

\begin{document}

\begin{textblock}{15}(0.5,14.7)
{
\noindent\hrulefill

\noindent\fontsize{8pt}{8pt}\selectfont\copyright\ 2022 IEEE. Personal use of this material is permitted. Permission from IEEE must be obtained for all other uses, in any current or future media, including reprinting/republishing this material for advertising or promotional purposes, creating new collective works, for resale or redistribution to servers or lists, or reuse of any copyrighted component of this work in other works. \hspace{5pt} This is the accepted version of: M. Sul\'ir, M. Regeci. Software Engineers' Questions and Answers on Stack Exchange. 2022 IEEE 16th International Scientific Conference on Informatics, IEEE, 2022, pp. 304--310. \url{http://doi.org/10.1109/Informatics57926.2022.10083403}

}
\end{textblock}

\title{Software Engineers' Questions and Answers\\on Stack Exchange}

\author{\IEEEauthorblockN{Mat\'u\v{s} Sul\'ir, Marcel Regeci}
\IEEEauthorblockA{\textit{Department of Computers and Informatics} \\
\textit{Faculty of Electrical Engineering and Informatics} \\
\textit{Technical University of Ko\v{s}ice}\\
Letná 9, 042 00 Ko\v{s}ice, Slovakia \\
matus.sulir@tuke.sk, marcel.regeci@student.tuke.sk}
}

\maketitle

\begin{abstract}
There exists a large number of research works analyzing questions and answers on the popular Stack Overflow website. However, other sub-sites of the Stack Exchange platform are studied rarely. In this paper, we analyze the questions and answers on the Software Engineering Stack Exchange site that encompasses a broader set of areas, such as testing or software processes. Topics and quantities of the questions, historical trends, and the authors' sentiment were analyzed using downloaded datasets. We found that the asked questions are most frequently related to database systems, quality assurance, and agile software development. The most attractive topics were career and teamwork problems, and the least attractive ones were network programming and software modeling. Historically, the topic of domain-driven design recorded the highest rise, and jobs and career the most significant fall. The number of new questions dropped, while the portion of unanswered ones increased.
\end{abstract}


\section{Introduction}

The Stack Exchange platform is a network of Q\&A (question and answer) sites that enable people to collaborate effectively in finding solutions and answers to questions related to a wide variety of topics. It is used by millions of people every day to solve problems they encounter in their jobs or learn more about their hobbies. The original, oldest site -- Stack Overflow\footnote{\url{https://stackoverflow.com}} -- is focused on programming questions related to specific source code excerpts, such as algorithmic problems in a given program or technical questions about frameworks and libraries. Over the course of years, many sub-sites emerged on the Stack Exchange platform, including sites focused on mathematics, physics, or even cooking. One of the components of the platform is the Software Engineering Stack Exchange\footnote{\url{https://softwareengineering.stackexchange.com}} site designed for professionals, academics, and students working in the software systems development lifecycle.

Software Engineering Stack Exchange is the 16th oldest site of the Stack Exchange
platform, with about 350,000 users and 60,000 asked questions\footnote{\url{https://stackexchange.com/sites?view=list\#technology-traffic}}. The focus of this site is
defined by a set of adequate question areas, which are namely: software development methods and procedures, requirements, architecture and design, quality assurance and testing, configuration management, assembly, release, and deployment\footnote{\url{https://softwareengineering.stackexchange.com/help/on-topic}}.

In the last decade, numerous papers analyzed Stack Overflow from various aspects. For example, Barua et al. \cite{Barua14what} provided a general overview of topics and trends. Asaduzzaman et al. \cite{Asaduzzaman13answering} focused on unanswered questions. Other papers analyzed specific topics of questions, such as chatbots \cite{Abdellatif20challenges} or concurrent programming \cite{Ahmed18what}.

However, we are not aware of any paper analyzing the Software Engineering Stack Exchange site, with the exception of a study by Verma et al. \cite{Verma19compehensive} that focused solely on the analysis of topics using LDA (Latent Dirichlet Allocation) during the years 2015--2017.

Therefore, in this paper, we would like to provide the results of the analysis of Software Engineering Stack Exchange from its inception in 2008 until April 2021, comprising the discussion of the most common and attractive topics, their historical development, the trends in the numbers of (un)answered questions asked over time, and the sentiment of the authors.

The analysis of Software Engineering Stack Exchange is interesting for two main reasons. First, due to the long-term operation of the platform and its high popularity, a large amount of data was collected, which constitutes a unique representation of the software engineers' interests and attitudes. Second, the analysis of the data could help better understand the platform and thus optimize its operation and increase the effectiveness of collective cooperation of users by developing and applying new methods and policies.


\section{Related Work}

A vast number of research works analyzed the main site of the Stack Exchange platform: Stack Overflow. Barua et al. \cite{Barua14what} provided an overview of the main discussion topics and their coupling, along with the changes in developers' interests and technologies used over time.

Asaduzzaman et al. \cite{Asaduzzaman13answering} mined unanswered questions from the mentioned website and tried to reveal characteristics that make questions difficult to answer. Baltes et al. \cite{Baltes20code} studied the duplication of code snippets that are often included on Stack Overflow. Zhang et al. \cite{Zhang21empirical} found that obsolete answers are sometimes posted on this website, and they are rarely updated.

Beyer et al. \cite{Beyer20what} tried to automatically classify questions on Stack Overflow into categories, such as API (application programming interface) usage, discrepancy, errors, review, or conceptual.

Some papers analyzed specific question topics on Stack Overflow: the challenges of chatbot development \cite{Abdellatif20challenges}, questions about concurrent programming \cite{Ahmed18what}, web developers' discussions \cite{Bajaj14mining}, privacy-related questions \cite{Tahaei20understanding}, and Docker challenges \cite{Haque20challenges}.

A few research works analyzed other sub-sites on the Stack Exchange platform. Among them, Kou and Gray \cite{Kou19practice} studied the evolution of UX (user experience) knowledge on UX Stack Exchange. Kamienski and Bezemer \cite{Kamienski21empirical} used multiple question\&answer websites for game developers, including Game Development Stack Exchange, as their data sources.

In contrast to the mentioned papers, we focused on the Software Engineering Stack Exchange website and its topic areas. Although a study by Verma et al. \cite{Verma19compehensive} also used the mentioned website as its data source, there are two key differences. First, they analyzed solely the topics and trends using LDA, while we included also research questions about (un)answered question quantities and the authors' sentiment in our study. Second, the study by Verma et al. focused only on the years 2015--2017, while we analyzed much broader year ranges.


\section{Research Questions}

We were interested in the following research questions:

\begin{description}
\item[RQ1] \textit{What are the most common topics of questions asked at Software Engineering Stack Exchange?}
\item[RQ2] \textit{Which topics attract the most and least attention?}
\item[RQ3] \textit{What was the historical development of the topics?}
\item[RQ4] \textit{How did the number of all and unanswered newly-asked questions change over time?}
\item[RQ5] \textit{Is the sentiment of the answers related to the reputation of their authors?}
\end{description}

In the following sections, we will always describe the methods used to answer each of the research questions and then present the results.


\section{Common Topics}
\label{s:frequent}

First, we would like to know to what topics the software engineers' questions on the platform most frequently pertain, which is the subject of RQ1.

\subsection{Method}

To obtain the data for analysis, we used the Stack Exchange Data Explorer\footnote{\url{https://data.stackexchange.com}}. Using an SQL query, we downloaded all questions that have a score of 1 or more (meaning they were marked as useful and relevant by at least one user) and have an accepted answer associated with them. We appended the text of the accepted answer to the body of the question since the author of the accepted answer is often an expert that uses proper domain terms, which should improve the quality of topic modeling. The resulting 28,077 questions were saved into a CSV (comma-separated values) file.

Then we pre-processed the text of the questions (including the appended accepted answers). Newline characters, punctuation, HTML tags, and code samples were removed. The text was transformed into a list of tokens -- words in lowercase and without any special characters. We removed stopwords, such as articles, pronouns, or prepositions. The words were lemmatized, i.e., transformed to a base form (e.g., solving into solve). Frequently co-occurring words were merged into single tokens representing bigrams, for instance, ``middle'' and ``ground'' into ``middle\_ground''.

From the list of tokens, we created a dictionary by filtering out tokens that were present in either less than 30 documents (questions) or more than 20\% of documents. This way, we excluded too specific or too general terms that would impair topic recognition. A corpus was then created, assigning the frequency of each token in the dictionary to every document.

The dictionary and the corpus were used as input of Mallet LDA \cite{Kachites02mallet}. The best model, based on the coherence score, was obtained for the number of topics set to 50. LDA assigns each document a set of topics along with a relevance score, which was used to determine the main (most relevant) topic for each question.

Since the topics produced by LDA are unnamed by themselves, we named each of the 50 topics by inspecting the list of the terms most relevant to it (as suggested by Mallet LDA) and manually reading a few associated documents if necessary.

\subsection{Results}

In Table~\ref{t:frequent}, we can see the list of the 20 most frequent topics of questions. The first column represents the portion of questions for which the given topic was identified as the main topic by LDA. The second column contains the name of the topic.

\begin{table}[h]
\renewcommand{\arraystretch}{1.3}
\caption{The Top 20 Most Frequent Topics of the Asked Questions}
\label{t:frequent}
\centering
\begin{tabular}{|r|l|}
\hline
\textbf{Frequency} & \textbf{Topic} \\ \hline
3.9 \% & database systems \\ \hline
3.7 \% & quality assurance \\ \hline
3.5 \% & agile software development \\ \hline
3.4 \% & software licensing \\ \hline
3.2 \% & asynchronous program execution \\ \hline
3.1 \% & jobs and career \\ \hline
3.0 \% & version control systems \\ \hline
2.8 \% & web application frontend \\ \hline
2.8 \% & encapsulation \\ \hline
2.8 \% & inheritance \\ \hline
2.7 \% & education and research \\ \hline
2.6 \% & domain-driven design \\ \hline
2.5 \% & software configuration \\ \hline
2.4 \% & functional programming \\ \hline
2.3 \% & recursion \\ \hline
2.3 \% & source code compilation \\ \hline
2.3 \% & authentication and authorization \\ \hline
2.3 \% & Model-View-Controller architecture \\ \hline
2.2 \% & exception handling \\ \hline
2.0 \% & network and backend \\ \hline
\end{tabular}
\end{table}

The asked questions were most frequently related to databases, quality assurance, agile software development, licensing, and asynchronous programming. However, the range of topics is quite diverse, and none of them prevails significantly.

The topic of database systems contained questions from a wide variety of sub-topics. Both traditional relational databases and NoSQL systems were discussed. The questions about relational databases included topics such as table joining, normalization and denormalization, indexes, views, or triggers. Many questions were related to performance concerns, particularly to the compromise between a clean database design and scalability. Object-relational mapping (ORM) was also frequently discussed. There were also questions looking at databases more broadly, e.g., about the design of an online database creator.

Questions about quality assurance were frequently related to test-driven development. Unit tests, integration tests, and particularly the distinction between them were also often discussed.

Agile software development questions discussed mainly terms related to Scrum (e.g., user stories, backlog, sprint planning, task estimation, story points), both from the theoretical point of view and specific practical problems that arose.


\section{Most and Least Attractive Topics}

Although questions regarding some topic may be frequently asked, it does not automatically mean this topic attracts attention from software engineers. For example, some questions are viewed by only a small group of visitors and remain unanswered for a long time, while others are frequently searched, viewed, and then heavily commented and answered. Therefore, in RQ2, we decided to study the attractiveness of individual topics.

\subsection{Method}

In RQ2, we used the data from the previous research question (RQ1), i.e., the complete Mallet LDA model. The names of all 50 topics and the assignment of each question to its main topic were retained.

For each question in the dataset, we calculated the Accumulated Post Score (AMS), which is a dimensionless number expressing the question's importance based on its selected properties, such as the upvote and comment count. It was defined by Bajaj et al. \cite{Bajaj14mining} as:
\[
AMS = 3 U - 25 D + 10 C + A + F
\]
In the formula, $U$ is the number of upvotes, $D$ the number of downvotes, $C$ is the comment count, $A$ the answer count, and $F$ the number of people who marked the question as favorite. In contrast to Bajaj et al. \cite{Bajaj14mining}, the numbers of upvotes and downvotes were not counted separately, but as a total score, i.e., either positive (upvotes) or negative (downvotes).

Then we calculated the average AMS for every topic as a sum of the scores of each question for which this was the main topic divided by the number of questions.

\subsection{Results}

All topics were sorted by their average AMS. In Table~\ref{t:most-attractive}, there is a list of the most attractive topics according to this metric. The top three most attractive topics were jobs and career, teamwork problems, and code readability.

We can conclude that ``jobs and career'' is both a very attractive and frequently asked topic. The second item on the list, teamwork problems, also represents a socially oriented topic, which is often controversial and thus tends to produce more elaborate discussions. Code readability questions were frequently related to concerns such as style, comments, and indentation, which traditionally cause disputes between the proponents of individual options.

\begin{table}[h]
\renewcommand{\arraystretch}{1.3}
\caption{The 10 Most Attractive Topics}
\label{t:most-attractive}
\centering
\begin{tabular}{|r|l|}
\hline
\textbf{Score} & \textbf{Topic} \\ \hline
69.6 & jobs and career \\ \hline
67.1 & teamwork problems \\ \hline
63.7 & code readability \\ \hline
61.9 & code refactoring \\ \hline
49.7 & education and research \\ \hline
48.6 & time management \\ \hline
48.4 & choosing programming language \\ \hline
47.0 & functional programming \\ \hline
45.9 & version control systems \\ \hline
44.1 & hardware and operating systems \\ \hline
\end{tabular}
\end{table}

A list of the least attractive topics with their corresponding AMS is displayed in Table~\ref{t:least-attractive}. Questions related to network programming and web application backend development, software modeling, and access control attract too little attention from the platform users after they are asked.

One possible reason for the unattractiveness of network programming and web backend development questions on this website might probably be the existence of not only a well-known separate Stack Exchange site for programming (Stack Overflow) but also a site for networking (Networking Stack Exchange). Software modeling and particularly model-driven software development is often considered an academic approach by practitioners \cite{Poruban15teaching}. Finally, access control questions might face a similar problem as was already mentioned: There exists a sub-site of Stack Exchange dedicated to information security.

\begin{table}[h]
\renewcommand{\arraystretch}{1.3}
\caption{The 10 Least Attractive Topics}
\label{t:least-attractive}
\centering
\begin{tabular}{|r|l|}
\hline
\textbf{Score} & \textbf{Topic} \\ \hline
16.8 & network and backend \\ \hline
19.3 & sofware modeling \\ \hline
19.3 & access control \\ \hline
19.4 & user data collection \\ \hline
20.6 & asynchronous program execution \\ \hline
21.0 & Model-View-Controller architecture \\ \hline
21.6 & recursion \\ \hline
22.0 & software development business contract \\ \hline
22.8 & web services and microservices \\ \hline
22.9 & sorting algorithms \\ \hline
\end{tabular}
\end{table}


\section{Historical Trends}

In RQ3, we will look at the development of the popularity of individual topics over the years, focusing on the most rising and falling trends.

\subsection{Method}

For each year, we calculated the relative frequency of every topic, using the LDA model data from the previous research questions. We divided the number of questions in a given year for which a given topic was the main one by the total number of questions in this year. Since the number of questions in the years 2008 and 2009 was too small to be relevant, and the year 2021 contained incomplete data, we included only the years 2010--2020 in the analysis.

Next, we marked the topics whose relative frequency was higher in 2020 than in 2010 as rising and the rest of them as falling. For both rising and falling trends, we found three topics with the highest difference between the maximum and minimum relative frequency during the analyzed years.

We also investigated the historical development of the three most frequent topics from section~\ref{s:frequent}.

\subsection{Results}

\begin{figure}[h]
\centering
\includegraphics[width=\linewidth]{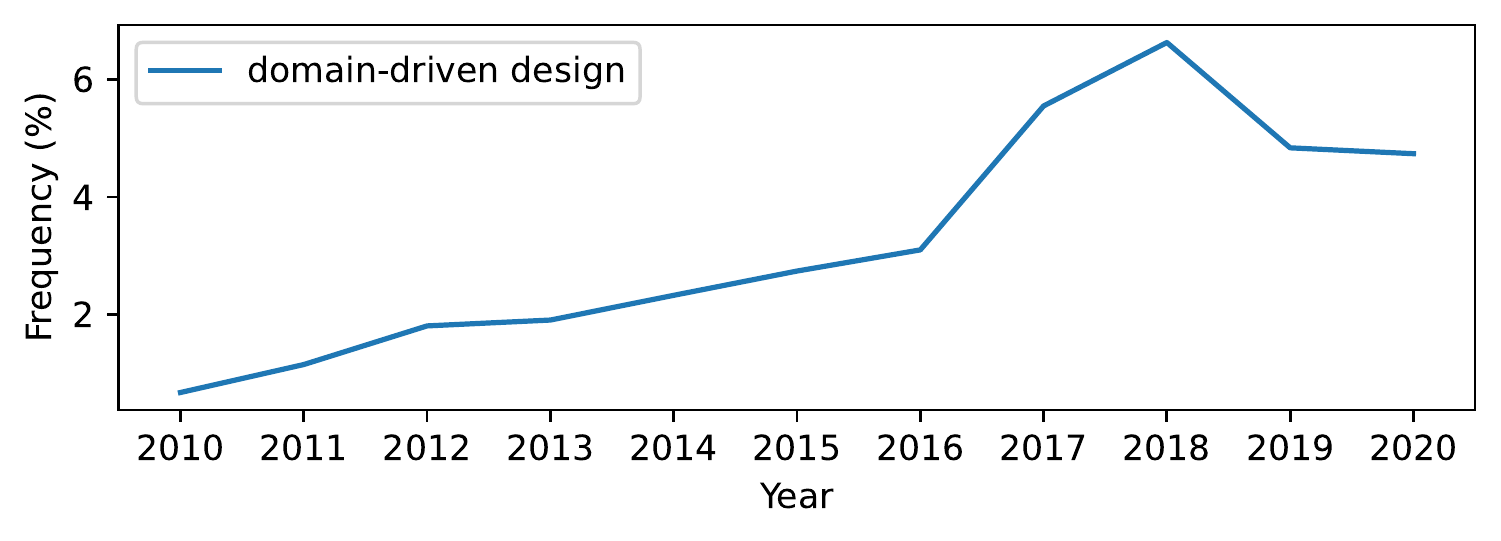}
\vspace{1mm}

\includegraphics[width=\linewidth]{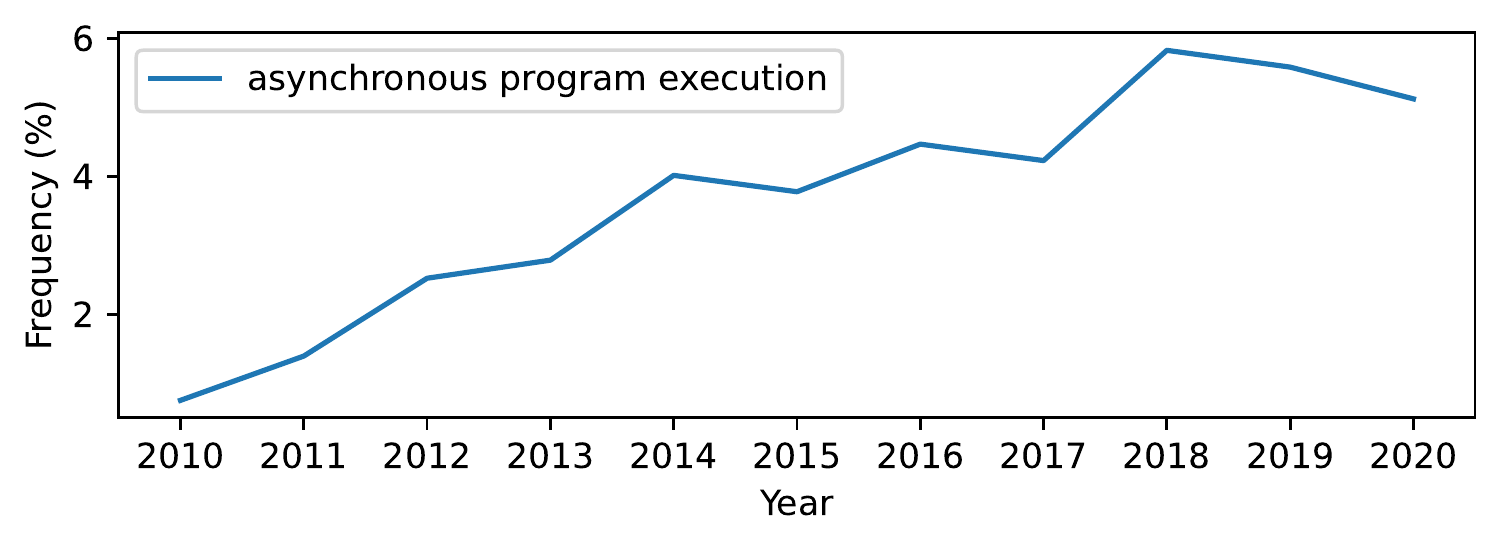}
\vspace{1mm}

\includegraphics[width=\linewidth]{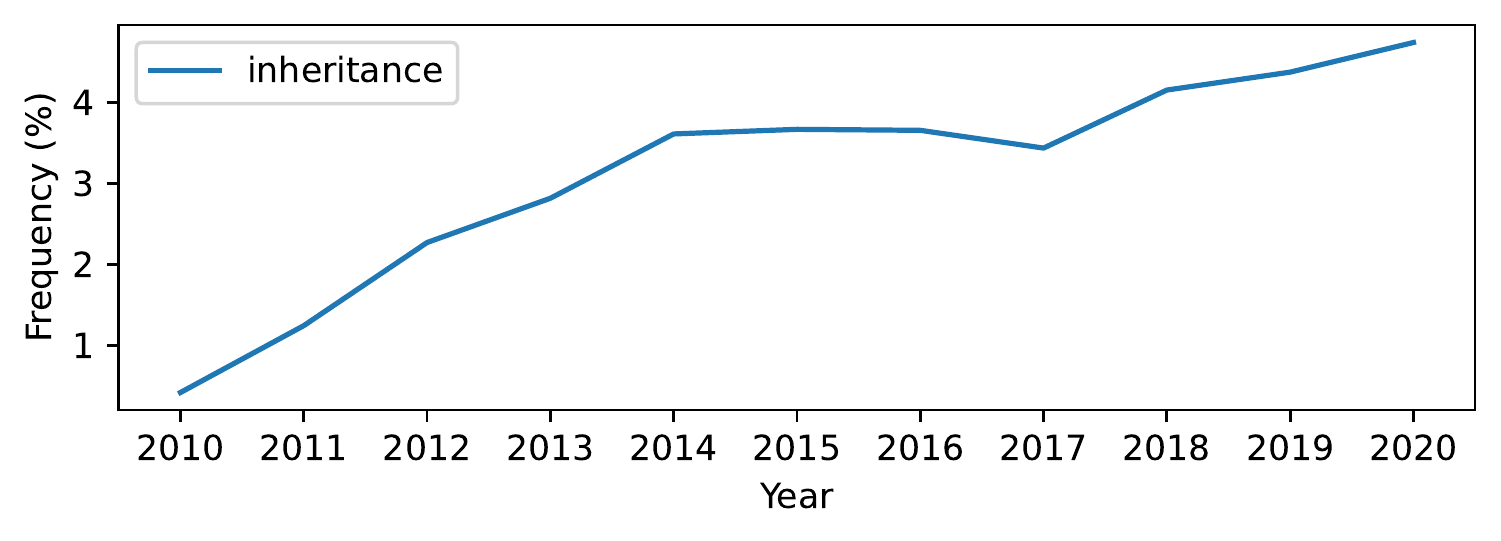}
\caption{The most rising topics.}
\label{f:rising}
\end{figure}

In Fig.~\ref{f:rising}, there are plots of the three most rising trends: domain-driven design, asynchronous programming, and inheritance. The relative frequency of the most rising topic, domain-driven development, increased every year until 2018 but started to slightly decrease in the following years. As the domain-driven design is now almost a two-decade-old concept, this might mean the discussion about it starts losing intensity. However, it could also be only a temporary fluctuation.

\begin{figure}[h]
\centering
\includegraphics[width=\linewidth]{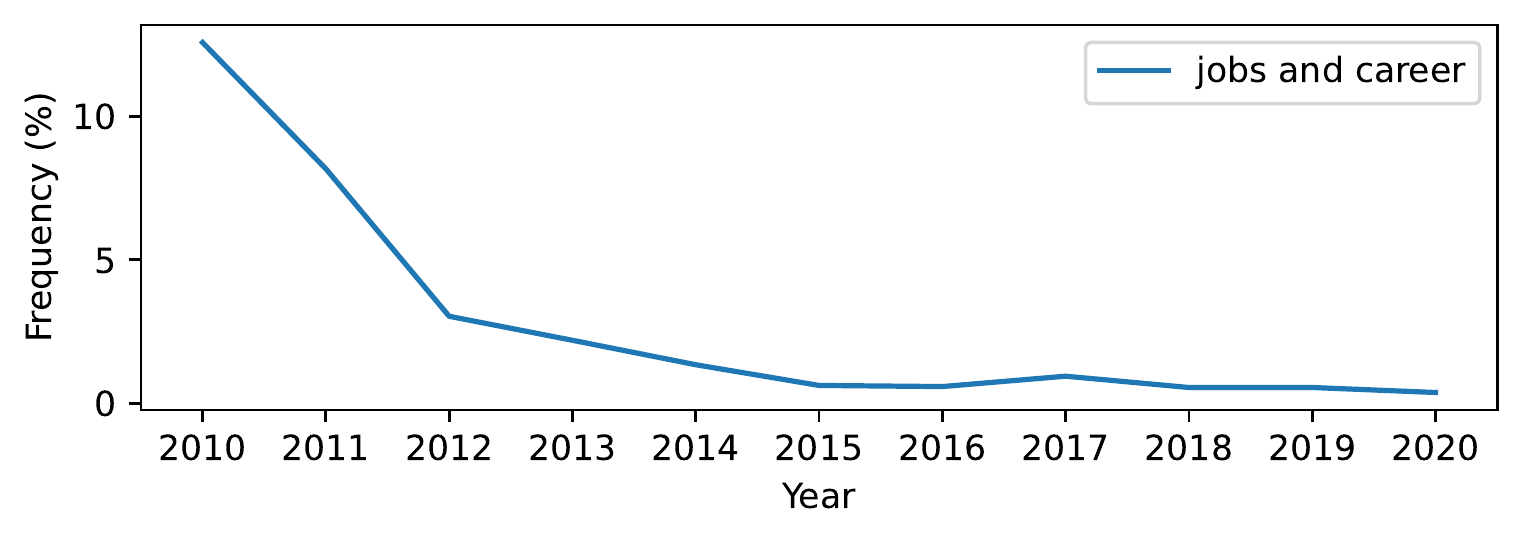}
\vspace{1mm}

\includegraphics[width=\linewidth]{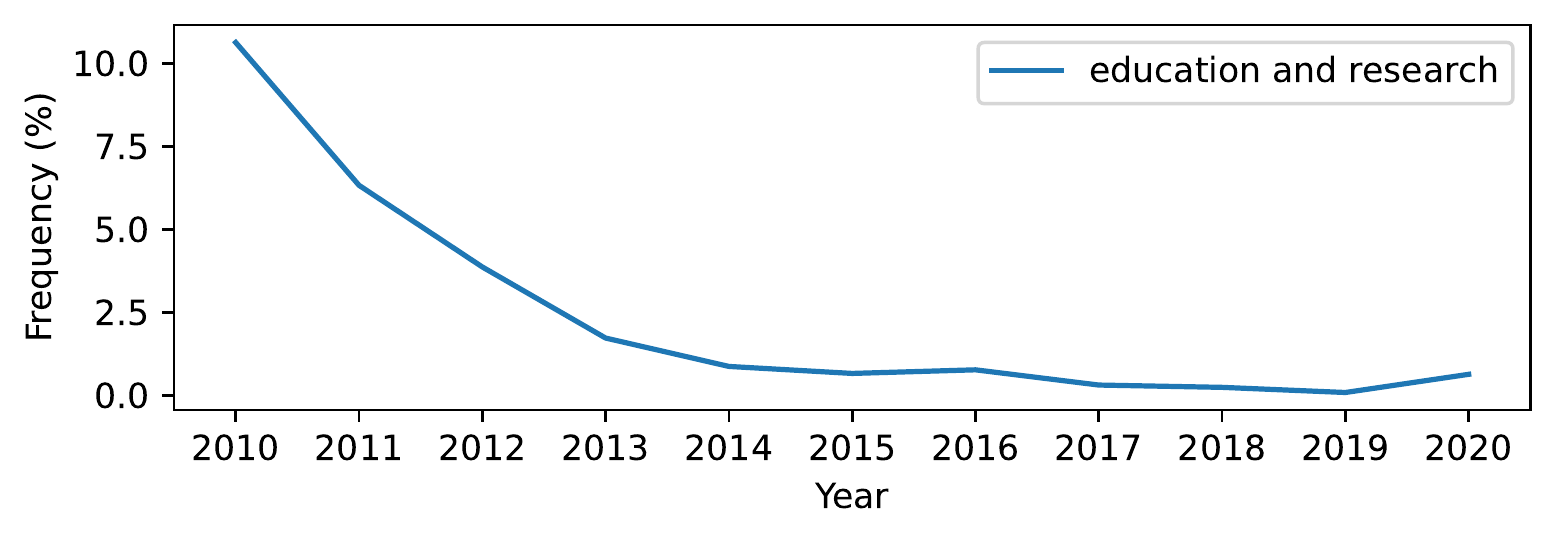}
\vspace{1mm}

\includegraphics[width=\linewidth]{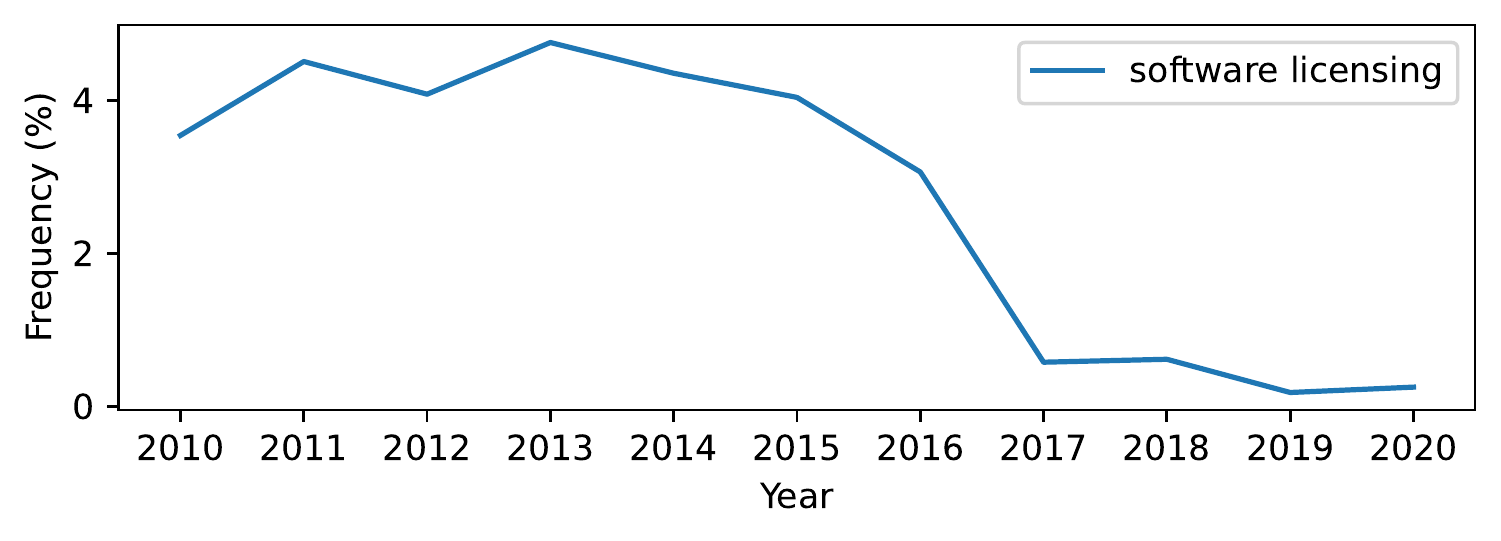}
\caption{The most falling topics.}
\label{f:falling}
\end{figure}

Fig.~\ref{f:falling} displays the plots of the three most falling trends: jobs and career, education and research, and software licensing.

For the first two mentioned topics, the relative frequency was extremely high in the first years -- in 2010 more than 10\% of questions concerned these topics. Then the frequency began to fall rapidly. We manually inspected a small sample of these questions, and we found that they are often closed as being off-topic. According to Software Engineering Stack Exchange rules, questions regarding career or education advice are considered off-topic because they are often meaningful only for a specific asker, and answers to them are almost always subjective. Licensing questions are also being closed since the website claims it does not aim to provide legal advice.

\begin{figure}[h]
\centering
\includegraphics[width=\linewidth]{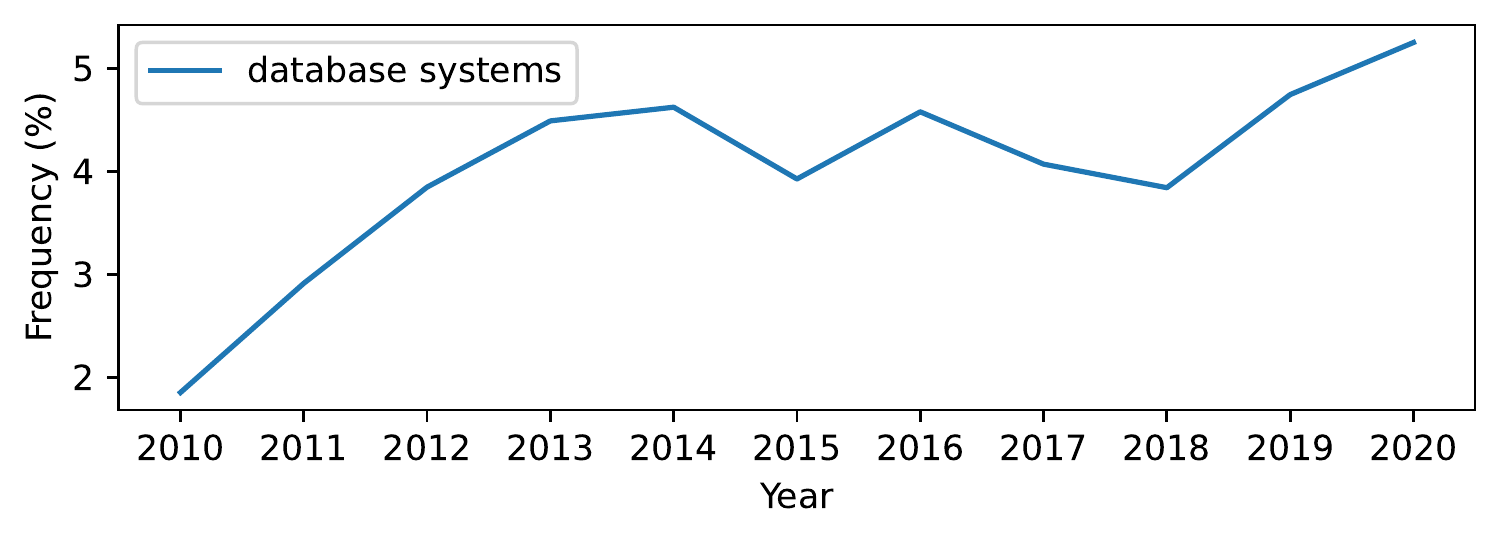}
\vspace{1mm}

\includegraphics[width=\linewidth]{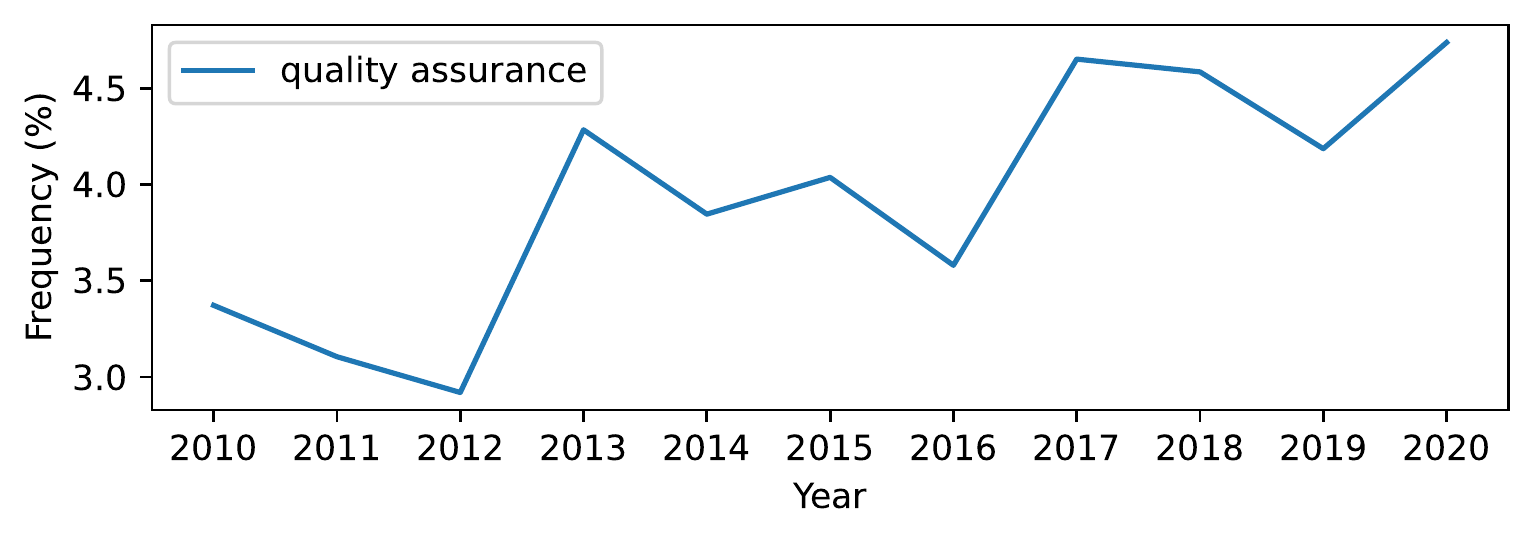}
\vspace{1mm}

\includegraphics[width=\linewidth]{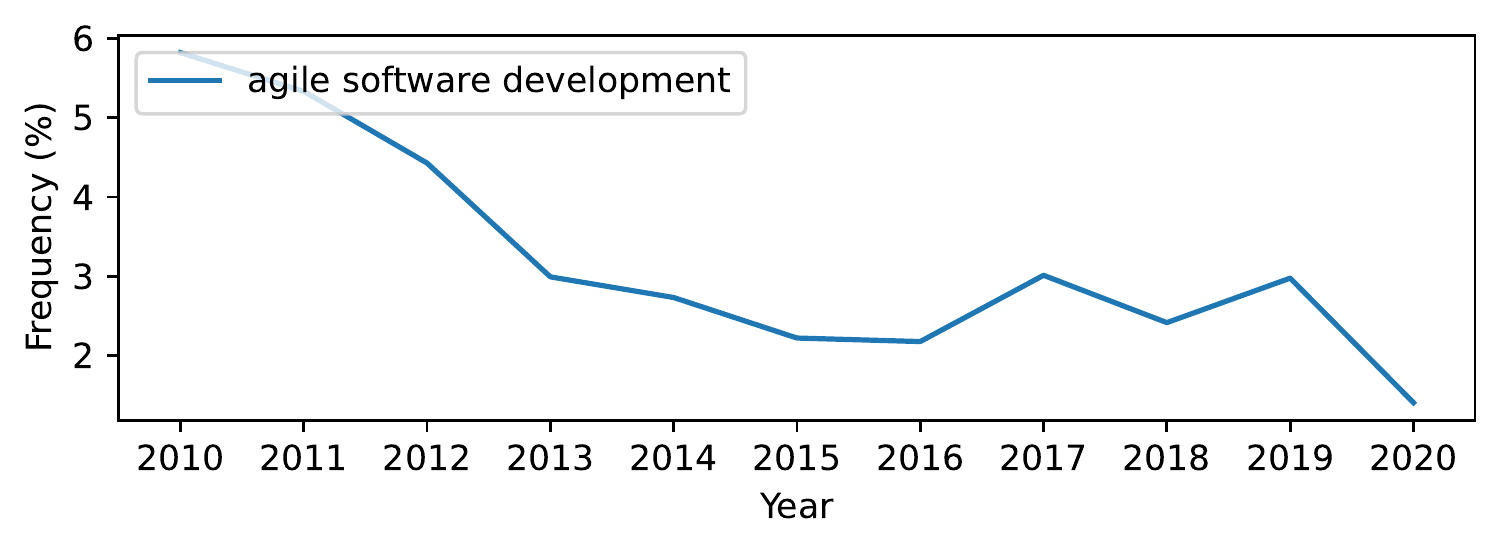}
\caption{The historical trends of the three most frequent topics.}
\label{f:frequent-history}
\end{figure}

Historical trends in the popularity of the three most frequent topics from section~\ref{s:frequent} are displayed in Fig.~\ref{f:frequent-history}. Database systems were rising until 2014, when they started to stagnate. The topic of quality assurance fluctuates, but it shows a generally positive trend. Finally, agile software development fell from about 6\% to less than 2\% over the decade.

\section{Question Count}

In RQ4, we would like to find whether the overall number of questions asked on Software Engineering Stack Exchange rises or falls over time. Furthermore, we will be interested in the historical development of the proportion of questions without an accepted answer. This will provide us insights about the health of the platform and its potential problems.

\subsection{Method}

Using a Stack Exchange Data Explorer query, we determined the total numbers of questions asked during each of the years from 2010 to 2020.

After a question is asked, its author should mark one of the answers as accepted in case of satisfaction, but in reality, many questions remain unresolved. Therefore, we also queried the counts of questions asked in the given years that do not have an accepted answer associated with them.

\subsection{Results}

\begin{figure*}[h]
\centering
\includegraphics[width=0.68\textwidth]{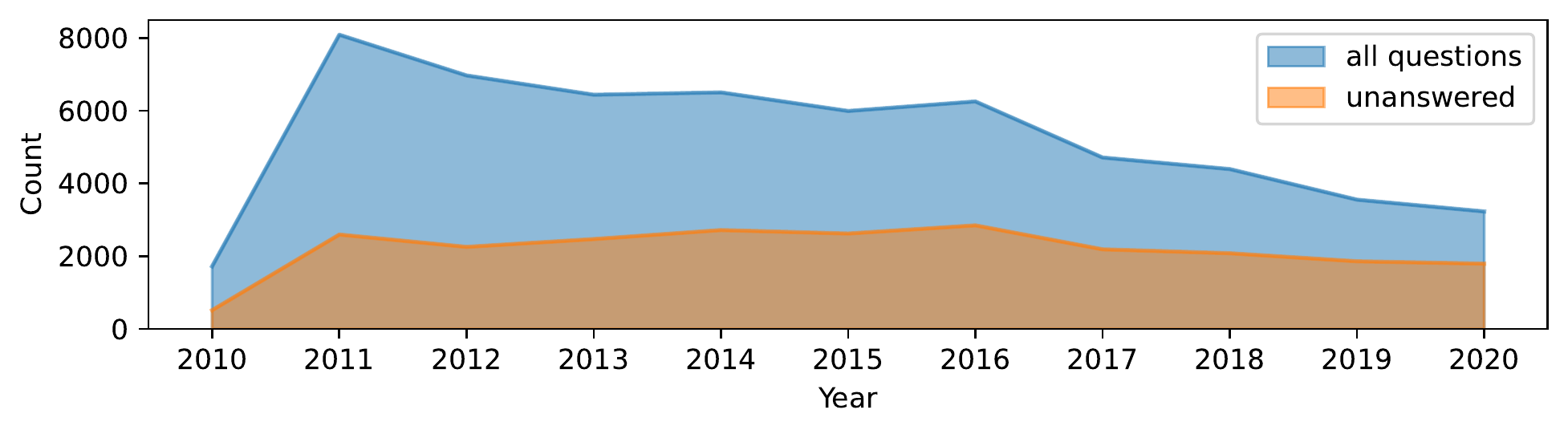}
\caption{The counts of all and unanswered questions asked in the given years.}
\label{f:questions}
\end{figure*}

The plot of the number of all and unanswered questions asked over time is shown in Fig.~\ref{f:questions}. The overall asked question count increased almost four-fold from 2010 to 2011. After this year it started to decline steadily.

An interesting observation is that the number of questions without an accepted answer remained almost constant during the years 2011--2020. This means the portion of unanswered questions increased dramatically during these nine years, specifically from 32\% to 56\%.


\section{Authors' Sentiment}

Ideally, the authors on the Stack Exchange platform should remain objective and avoid too emotional responses. In reality, they sometimes express also emotions in their posts. In RQ5 we would like to determine if there is a relationship between the sentiment (subjectivity and polarity) of the answers and the reputation of their authors. The reputation on Stack Exchange is a numerical score that the users earn for asking, answering, and editing questions. Basically, a higher reputation means higher experience with the given site.

\subsection{Method}

Using Stack Exchange Data Explorer, we downloaded the texts of answers having at least 100 characters since shorter text could be unsuitable for sentiment analysis. To prevent the assessment of users based on merely one or two posts, only users having at least 3 such answers on the website were included in our query. Because of the limitations of the tool, for users having more than fifteen answers we selected a random sample of size 15. The resulting dataset contained 49,284 answers.

Each answer was pre-processed by removing newlines, HTML tags, and code snippets. Polarity (positivity vs. negativity) and subjectivity (the degree to which the given text is subjective) were computed using the TextBlob library \cite{Loria20textblob}.

For every user, we calculated the average subjectivity and polarity. A measure of correlation capturing also nonlinear relationships, Spearman's rho, was computed first between the reputation and subjectivity and then between the reputation and polarity of users.

\subsection{Results}

\begin{figure}[h]
\centering
\includegraphics[width=\linewidth]{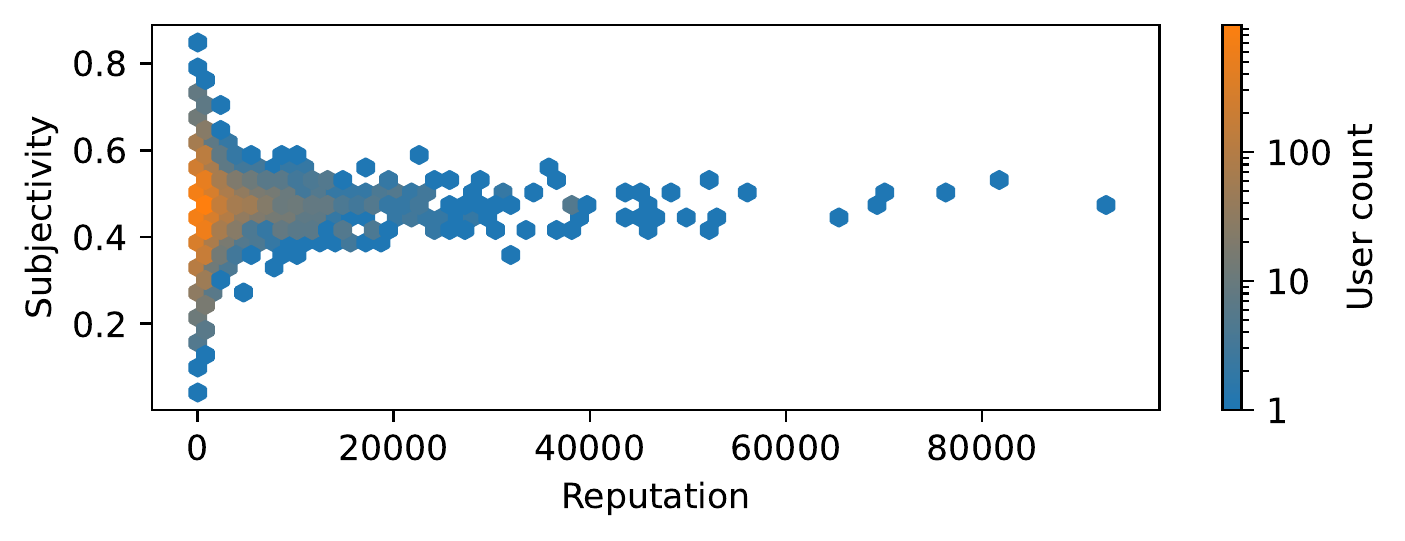}
\caption{The relationship between the users' reputation and the average subjectivity of their answers.}
\label{f:subjectivity}
\end{figure}

Fig.~\ref{f:subjectivity} displays a hexbin plot, where the reputation of users is on the x-axis and the average subjectivity on the y-axis. The logarithmic color scale represents the number of users in each hexagonal bin. Extreme outliers, namely 8 users with a reputation higher than 100,000, were removed from the plot. While there exist some predominantly subjective or objective users, a vast majority is located in the middle. The plot shows no signs of correlation, which is confirmed by Spearman's rho equal to 0.05 (where 0 means no correlation and 1 perfect positive correlation).

\begin{figure}[h]
\centering
\includegraphics[width=\linewidth]{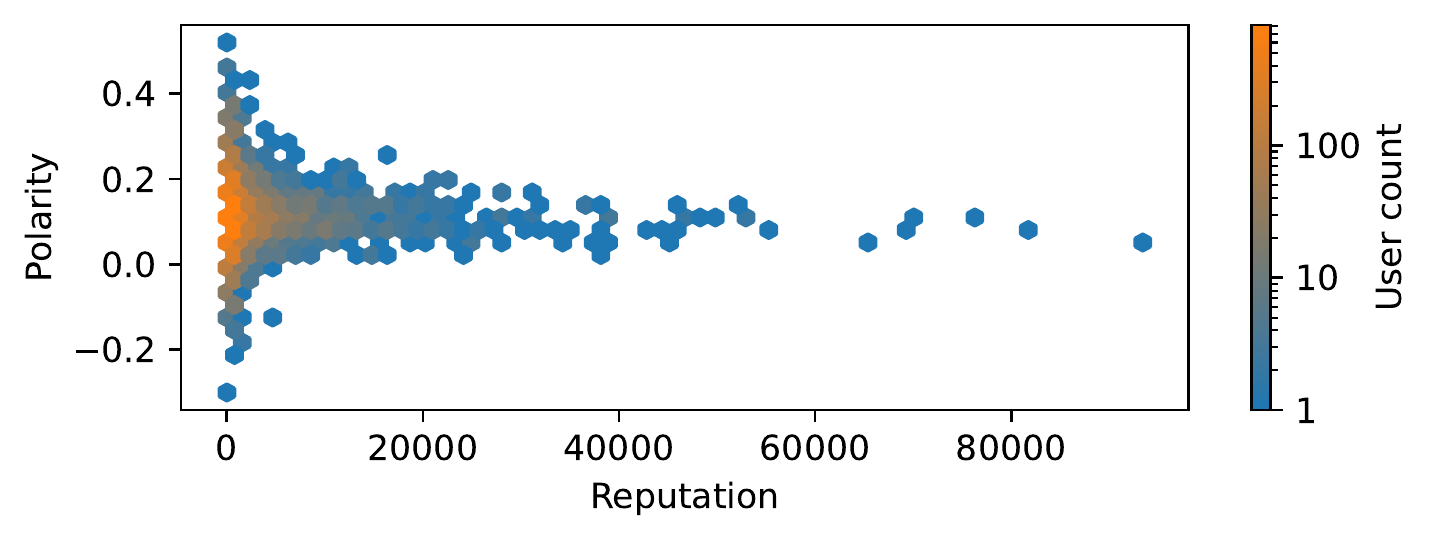}
\caption{The relationship between the users' reputation and the average polarity of their answers.}
\label{f:polarity}
\end{figure}

Similarly, Fig.~\ref{f:polarity} contains a hexbin plot of the users' average polarity based on their reputation. The majority of users tend to be neither too positive nor too negative, but the average inclines slightly towards positivity. The computed Spearman's rank correlation coefficient is -0.01, i.e., there is no correlation between reputation and polarity.

\section{Threats to Validity}

Now we will describe threats to the validity of our study.

\subsection{Construct Validity}

For the individual research questions, we analyzed only subsets of all data available on Software Engineering Stack Exchange. In RQ1--RQ3, we used only questions having a positive score and an accepted answer. A limited year range was analyzed in RQ3 and RQ4, and a random subset of answers was queried for users with too many posts in RQ5. However, we always provided a rationale for the given selection criteria.

\subsection{Internal Validity}

Rapid changes in the historical development of topics might be caused by new or changed rules coming into effect. For example, the decline of the topic ``jobs and career'' was probably caused not by the lack of software engineers' interest in their careers but due to many questions about jobs being marked as not relevant by the community.

\subsection{External Validity}

The study was based on a single data source, namely the Software Engineering Stack Exchange site, so it might not be representative of all software engineers' questions about the software systems development lifecycle and related areas. Although the analyzed website contains more than 200,000 posts, extending the study to use multiple data sources would improve its external validity.

\subsection{Reliability}

The naming of individual LDA topics was performed by one of the authors. However, the other author checked all assignments based on the most relevant terms and also a subset of names by skimming the text of selected posts assigned to a given topic.

Jupyter notebooks and data dumps used to produce the results described in this study are publicly available in a permanent repository.\footnote{\url{https://doi.org/10.17605/OSF.IO/M8STH}}


\section{Conclusion and Future Work}

We analyzed the questions and answers on the Software Engineering Stack Exchange website from various aspects. According to our findings, the topics of most frequently asked questions were  database systems, quality assurance, and agile software development.

The most attractive topics, taking into account metrics such as upvote, answer, and comment counts, were jobs and career, teamwork problems, and code readability. On the other hand, questions related to web application backend, software modeling, and access control were among the least attractive ones.

The new question count on the analyzed website was rising until 2011 but started to decline gradually after this year. Since the number of unanswered questions remained almost the same, their relative proportion significantly increased.

During the years 2010--2020, domain-driven design, asynchronous programming, and inheritance were the most rising trends. Jobs and career, education and research, and software licensing question counts dropped the most. However, this does not mean software engineers are no longer interested in these questions -- instead, they started to be considered off-topic on the given website. This creates a fragmentation problem, where the developer must carefully consider at which Stack Exchange sub-site to ask a given question, if ever.

Regarding the authors' sentiment, we found no correlation between the reputation of users and the average subjectivity or polarity of their answers.

In the future, we could answer more research questions, particularly related to comments and bounties. We would also analyze more data sources, including the main Stack Overflow website and third-party question\&answer sites, and compare the results. Finally, other topical sub-sites of the Stack Exchange platform also provide a wealth of information suitable for analysis.

\section*{Acknowledgment}

This work was supported by VEGA Grant No.~1/0630/22 Lowering Programmers' Cognitive Load Using Context-Dependent Dialogs.

\balance

\bibliographystyle{IEEEtran}
\bibliography{informatics}

\begin{thebibliography}{10}
\providecommand{\url}[1]{#1}
\csname url@samestyle\endcsname
\providecommand{\newblock}{\relax}
\providecommand{\bibinfo}[2]{#2}
\providecommand{\BIBentrySTDinterwordspacing}{\spaceskip=0pt\relax}
\providecommand{\BIBentryALTinterwordstretchfactor}{4}
\providecommand{\BIBentryALTinterwordspacing}{\spaceskip=\fontdimen2\font plus
\BIBentryALTinterwordstretchfactor\fontdimen3\font minus
  \fontdimen4\font\relax}
\providecommand{\BIBforeignlanguage}[2]{{%
\expandafter\ifx\csname l@#1\endcsname\relax
\typeout{** WARNING: IEEEtran.bst: No hyphenation pattern has been}%
\typeout{** loaded for the language `#1'. Using the pattern for}%
\typeout{** the default language instead.}%
\else
\language=\csname l@#1\endcsname
\fi
#2}}
\providecommand{\BIBdecl}{\relax}
\BIBdecl

\bibitem{Barua14what}
A.~Barua, S.~W. Thomas, and A.~E. Hassan, ``\BIBforeignlanguage{English}{What
  are developers talking about? {A}n analysis of topics and trends in {Stack
  Overflow}},'' \emph{\BIBforeignlanguage{English}{Empirical Software
  Engineering}}, vol.~19, no.~3, pp. 619--654, 2014.

\bibitem{Asaduzzaman13answering}
M.~Asaduzzaman, A.~S. Mashiyat, C.~K. Roy, and K.~A. Schneider, ``Answering
  questions about unanswered questions of {S}tack {O}verflow,'' in \emph{2013
  10th Working Conference on Mining Software Repositories (MSR)}, 2013, pp.
  97--100.

\bibitem{Abdellatif20challenges}
A.~Abdellatif, D.~Costa, K.~Badran, R.~Abdalkareem, and E.~Shihab, ``Challenges
  in chatbot development: A study of {S}tack {O}verflow posts,'' ser. MSR
  '20.\hskip 1em plus 0.5em minus 0.4em\relax New York, NY, USA: Association
  for Computing Machinery, 2020, pp. 174--185.

\bibitem{Ahmed18what}
S.~Ahmed and M.~Bagherzadeh, ``What do concurrency developers ask about? {A}
  large-scale study using {S}tack {O}verflow,'' in \emph{Proceedings of the
  12th ACM/IEEE International Symposium on Empirical Software Engineering and
  Measurement}, ser. ESEM '18.\hskip 1em plus 0.5em minus 0.4em\relax New York,
  NY, USA: Association for Computing Machinery, 2018.

\bibitem{Verma19compehensive}
A.~Verma, N.~Sardana, and S.~Lal, ``Comprehensive analysis of trends in
  software engineering {Q\&A} site,'' in \emph{2019 9th International
  Conference on Cloud Computing, Data Science \& Engineering (Confluence)},
  2019, pp. 648--653.

\bibitem{Baltes20code}
S.~Baltes and C.~Treude, ``Code duplication on {S}tack {O}verflow,'' in
  \emph{Proceedings of the ACM/IEEE 42nd International Conference on Software
  Engineering: New Ideas and Emerging Results}, ser. ICSE-NIER '20.\hskip 1em
  plus 0.5em minus 0.4em\relax New York, NY, USA: Association for Computing
  Machinery, 2020, pp. 13--16.

\bibitem{Zhang21empirical}
H.~Zhang, S.~Wang, T.-H. Chen, Y.~Zou, and A.~E. Hassan, ``An empirical study
  of obsolete answers on {S}tack {O}verflow,'' \emph{IEEE Transactions on
  Software Engineering}, vol.~47, no.~4, pp. 850--862, 2021.

\bibitem{Beyer20what}
S.~Beyer, C.~Macho, M.~Di~Penta, and M.~Pinzger, ``What kind of questions do
  developers ask on {S}tack {O}verflow? {A} comparison of automated approaches
  to classify posts into question categories,'' \emph{Empirical Software
  Engineering}, vol.~25, no.~3, pp. 2258--2301, May 2020.

\bibitem{Bajaj14mining}
K.~Bajaj, K.~Pattabiraman, and A.~Mesbah, ``Mining questions asked by web
  developers,'' in \emph{Proceedings of the 11th Working Conference on Mining
  Software Repositories}, ser. MSR 2014.\hskip 1em plus 0.5em minus 0.4em\relax
  New York, NY, USA: Association for Computing Machinery, 2014, pp.
  112–--121.

\bibitem{Tahaei20understanding}
M.~Tahaei, K.~Vaniea, and N.~Saphra, ``Understanding privacy-related questions
  on {S}tack {O}verflow,'' ser. CHI '20.\hskip 1em plus 0.5em minus 0.4em\relax
  New York, NY, USA: Association for Computing Machinery, 2020, pp. 1--14.

\bibitem{Haque20challenges}
M.~U. Haque, L.~H. Iwaya, and M.~A. Babar, ``Challenges in {D}ocker
  development: A large-scale study using {S}tack {O}verflow,'' in
  \emph{Proceedings of the 14th ACM / IEEE International Symposium on Empirical
  Software Engineering and Measurement (ESEM)}, ser. ESEM '20.\hskip 1em plus
  0.5em minus 0.4em\relax New York, NY, USA: Association for Computing
  Machinery, 2020.

\bibitem{Kou19practice}
Y.~Kou and C.~M. Gray, ``A practice-led account of the conceptual evolution of
  {UX} knowledge,'' in \emph{Proceedings of the 2019 CHI Conference on Human
  Factors in Computing Systems}, ser. CHI '19.\hskip 1em plus 0.5em minus
  0.4em\relax New York, NY, USA: Association for Computing Machinery, 2019, pp.
  1--13.

\bibitem{Kamienski21empirical}
A.~Kamienski and C.-P. Bezemer, ``An empirical study of {Q\&A} websites for
  game developers,'' \emph{Empirical Software Engineering}, vol.~26, no.~6, pp.
  1--39, 2021.

\bibitem{Kachites02mallet}
A.~K. McCallum, ``M{ALLET}: A machine learning for language toolkit,'' 2002,
  http://mallet.cs.umass.edu.

\bibitem{Poruban15teaching}
J.~Porub\"an, M.~Ba\v{c}\'ikov\'a, S.~Chodarev, and M.~Nos\'a\v{l}, ``Teaching
  pragmatic model-driven software development,'' \emph{Computer Science and
  Information Systems}, vol.~12, no.~2, pp. 683--705, 2015.

\bibitem{Loria20textblob}
\BIBentryALTinterwordspacing
S.~Loria \emph{et~al.}, ``Text{B}lob, version 0.16.0,'' 2020. [Online].
  Available: \url{https://textblob.readthedocs.io}
\BIBentrySTDinterwordspacing

\end{thebibliography}

\end{document}